\documentclass[twocolumn]{aastex631}

\shorttitle{Red Asymmetry of H$_{\alpha}$ Line Profiles during the flares on II~Peg}
\shortauthors{Cao \& Gu}
\graphicspath{{./}{figures/}}

\begin{document}

\title{Red Asymmetry of H$_{\alpha}$ Line Profiles during the flares on the active RS~CVn-type star II~Pegasi}

\correspondingauthor{Dongtao Cao, Shenghong Gu}
\email{dtcao@ynao.ac.cn, shenghonggu@ynao.ac.cn}

\author[0000-0002-3534-1740]{Dongtao Cao}
\affiliation{Yunnan Observatories, Chinese Academy of Sciences, Kunming 650216, China}
\affiliation{Key Laboratory for the Structure and Evolution of Celestial Objects, Chinese Academy of Sciences, Kunming 650216, China}
\affiliation{International Centre of Supernovae, Yunnan Key Laboratory, Kunming 650216, China}

\author{Shenghong Gu}
\affiliation{Yunnan Observatories, Chinese Academy of Sciences, Kunming 650216, China}
\affiliation{Key Laboratory for the Structure and Evolution of Celestial Objects, Chinese Academy of Sciences, Kunming 650216, China}
\affiliation{School of Astronomy and Space Science, University of Chinese Academy of Sciences, Beijing 101408, China}



\begin{abstract}

Stellar coronal mass ejections (CMEs) have recently attracted much attention for their impacts on stellar evolution and surrounding exoplanets. RS~CVn-type stars could produce large flares, and therefore may have frequent CMEs. Here we report the capture of a possible CME or chromospheric condensation on the RS CVn-type star II~Pegasi (II~Peg) using high-resolution spectroscopic observation. Two flares were detected during the observation, and the low limits of the flare energies are of the order of $10^{33}$~erg and $10^{34}$~erg, respectively. Using mean spectrum subtraction, the H$_{\alpha}$ residual shows red asymmetry during the flares, and the redshifted broad emission components are probably caused by chromospheric condensation or coronal rain. Moreover, a far redshifted extra emission component with a high bulk velocity of 429~km~s$^{-1}$ was observed during the second flare and is probably due to a prominence eruption. The velocity greatly exceeds the star's escape velocity, which means that this eruption can develop into a CME. The CME mass is estimated to be 0.83--1.48~$\times~10^{20}$~g, which is slightly larger than the value expected from solar flare-CME extrapolation. The kinetic energy of CME, derived to be 0.76--1.15~$\times~10^{35}$~erg, is less than the kinetic energy extrapolated from solar events. Additionally, we could not completely rule out the possibility of chromospheric condensation resulting in the far redshifted extra emission. Finally, there is a blueshifted broad component in the subtracted H$_{\alpha}$ profile derived using synthesized spectral subtraction when no flare happened, and its behavior is associated with the H$_{\alpha}$ activity features.

\end{abstract}


\section{Introduction} \label{sec1}
Starspots, plages, flares, and prominences have been widely observed in cool stars \citep{Schrijver2000}. It is commonly accepted that all of these active phenomena arise from a powerful magnetic dynamo generated by the interplay between the turbulent motion in the convection zone and the stellar differential rotation, in a manner similar to the solar case. In recent years, stellar coronal mass ejections (CMEs) have attracted much attention, though they are more difficult to observe with current instrumentation because of having no spatial resolution. Based on some indirect evidences, such as Doppler-shifted emission or absorption signature in the optical, UV and X-ray spectral lines \citep[e.g.,][]{Houdebine1990, Leitzinger2011,  Argiroffi2019, Namekata2022, Inoue2023}, X-ray, extreme-UV (EUV), and far-UV (FUV) dimming \citep[e.g.,][]{Veronig2021, Namekata2023}, and X-ray continuous absorption \citep[e.g.,][]{Favata1999, Moschou2017}, researchers have made a lot of attempts to detect stellar CME events. Frequently occurred CMEs may be an important contribution to mass and angular momentum loss in the course of stellar evolution \citep[e.g.,][]{Aarnio2012, Osten2015}, and have severe impacts on the atmospheres of surrounding exoplanets \citep[e.g.,][]{Airapetian2016, Cherenkov2017, Hazra2022}.

RS CVn-type binary system includes at least one cool component showing particularly intense magnetic activity in several forms \citep{Hall1976, Fekel1986}. For example, prominence-like features have been detected in several RS CVn-type stars \citep[e.g.,][]{Hall1990, Hall1992, Cao2019, Cao2020}. Therefore, prominences may be common phenomena in this type stars. Moreover, RS CVn-type star has been frequently observed to produce strong flares and even superflares ($\ga~10^{33}$ erg) \citep[e.g.,][]{Tsuboi2016, Inoue2023}. We detected a series of possibly associated magnetic activity phenomena in a short time, including flare-related prominence activation, a strong optical flare, and post-flare loops, on RS~CVn-type star SZ~Psc \citep{Cao2019}. Based on the well-defined connection between highly energetic flares and CMEs on the Sun and the close relation between CMEs and eruptive prominences, it can be hypothesized that RS~CVn-type stars may have frequent CME occurrences. Here we focus on one of the most active and extensively studied single-lined RS~CVn-type binary stars, II~Peg.

The system II~Peg (=~HD~224085) consists of a K2~IV primary star and an unseen companion in an almost circular orbit with a period of about 6.7 days. The basic physical parameters of the K2~IV primary star are summarized in Table~\ref{tab1}. The K2~IV star shows remarkable photometric variability caused by large starspots \citep[e.g.,][]{Vogt1981, Lindborg2013}, $\mbox{Ca~{\sc ii}}$~H\&K and H$_{\alpha}$ line emission \citep[e.g.,][]{Vogt1981, Huenemoerder1987, Montes1997, Berdyugina1999, Frasca2008}, and UV and X-ray radiation \citep[e.g.,][]{Byrne1989, Ercolano2008}. Moreover, II~Peg belongs to stars to which Doppler imaging (DI) and Zeeman-Doppler imaging (ZDI) techniques were widely applied \citep[e.g.,][]{Gu2003, Lindborg2011, Hackman2012, Kochukhov2013, Xiang2014, Rosen2015, Strassmeier2019}.
\begin{deluxetable}{lll}
\tablenum{1}
\tablecaption{Basic Parameters of the K2~IV Primary Star of II~Peg\label{tab1}}
\tablewidth{0pt}
\tablehead{
\colhead{\bf{Parameter}} & \colhead{\bf{Value}} & \colhead{\bf{Reference}}
}
\startdata
Spectral Type             & K2~IV                  & \citet{Strassmeier1993}\\ 
$V_{max}$ (mag)        & 7.2                       & \citet{Strassmeier1993}\\
$d$ (pc)                      & 39.12 $\pm$0.08 & \citet{Gaia2020}\\
$P_{orb}$ (days)        & 6.7242078           & \citet{Rosen2015}\\
$P_{rot}$ (days)         & 6.7242078           & \citet{Rosen2015}\\ 
$T_{eff}$ (K)              & 4750~K                & \citet{Rosen2015}\\
$log~g$ (cgs)             & 3.5                       & \citet{Rosen2015}\\ 
$vsini$ (km~s$^{-1}$) & 21.6                     & This work \\
$M$ (M$_{\sun}$)      & 0.8                       & \citet{Berdyugina1998} \\
$R$ (R$_{\sun}$)       & 3.4                       & \citet{Berdyugina1998}\\
\enddata
\tablecomments{$V_{max}$ is maximum brightness in V-band, $d$ is stellar distance, $P_{orb}$ is orbital period, $P_{rot}$ is rotation period, $T_{eff}$ is effective temperature, $log~g$ is Log gravity, $vsini$ is projected rotational velocity, $M$ is mass, and $R$ is radius.}
\end{deluxetable}

II~Peg is a high-rate flaring star. Flares have been reported several times over a wide range of wavelength regions, such as optical, UV, and X-ray bands \citep[e.g.,][]{Rodono1987, Doyle1991, Berdyugina1999, Frasca2008, Ercolano2008, Siwak2010, Tsuboi2016}. In the optical spectral lines, sudden enhancements of the H$_{\alpha}$ and $\mbox{Ca~{\sc ii}}$ lines, as well as the emission feature in the $\mbox{He~{\sc i}}$ D$_{3}$ line, were usually attributed to flare events. For example, \citet{Berdyugina1999} detected two strong optical flares using the $\mbox{He~{\sc i}}$ D$_{3}$, $\mbox{Ca~{\sc ii}}$ K, and $\mbox{Ca~{\sc ii}}$ $\lambda$8498 lines. Broad blueshifted emissions could be found in the $\mbox{He~{\sc i}}$ D$_{3}$ and $\mbox{Ca~{\sc ii}}$ K line profiles at the flare maximum, which were attributed to the process of the explosive evaporation from the low chromosphere.

In this paper, we present the results about strong optical flares and flare-associated activity phenomena on II~Peg. In Section~\ref{sec2}, we provide details of our high-resolution spectroscopic observation and data reduction. The analysis of chromospheric activity indicators covered in observed spectra is described in Section~\ref{sec3}. In Section~\ref{sec4}, the behavior of chromospheric activity indicators and the associated magnetic activity phenomena of II~Peg during our observation are discussed in more detail. Finally, we give a summary and conclusions in Section~\ref{sec5}.

\section{Spectroscopic observation and data reduction}\label{sec2}
Spectroscopic data of II~Peg were obtained over nine nights from 2016 January~22 to 31. The observation was carried out with the fiber-fed high-resolution spectrograph (HRS), installed on the 2.16~m telescope at the Xinglong station of the National Astronomical Observatories, Chinese Academy of Sciences, China \citep{Fan2016}. HRS produces spectra with a resolving power R~=~$\lambda$/$\Delta\lambda$~$\simeq$~48000 in a wavelength range of 3900--10000~\AA, using a $4096 \times 4096$ pixel CCD detector.

Table~\ref{tab2} lists the information of our observation, which includes the observing date, exposure time, heliocentric Julian date (HJD), and orbital phase calculated with the ephemeris:
\begin{eqnarray}
HJD = 2,448,942.428 + 6^{d}.7242078 \times E
\end{eqnarray}
from \citet{Rosen2015}, where the zero-point is a time of the K2~IV primary star of II~Peg having maximum positive radial velocity. In addition, we observed some slowly rotating, inactive stars with the same spectral type and luminosity class as II~Peg. They are required for the spectral subtraction technique described in Section 3.1.

Data reduction was performed with the IRAF\footnote{IRAF is distributed by the National Optical Astronomy Observatories, which is operated by the Association of Universities for Research in Astronomy (AURA), Inc., under cooperative agreement with the National Science Foundation.} package, following the standard procedures (image trimming, bias correction, flat-field division, scattered light subtraction, one-dimensional spectrum extraction, and wavelength calibration). The wavelength calibration was obtained by using emission lines of Th-Ar spectra of the corresponding nights. Finally, all spectra were normalized by using low-order polynomial fits to the observed continuums with the CONTINUUM task in the IRAF package.

Examples of the normalized $\mbox{Ca~{\sc ii}}$ IRT ($\lambda$8662, $\lambda$8542, and $\lambda$8498), $\mbox{He~{\sc i}}$~$\lambda$6678, H$_{\alpha}$, $\mbox{Na~{\sc i}}$ D$_{1}$, D$_{2}$ doublet, $\mbox{He~{\sc i}}$ D$_{3}$, and H$_{\beta}$ line profiles of II~Peg are displayed in Figure~\ref{Fig1}.
\begin{deluxetable}{cccc}
\tablenum{2}
\tablecaption{II~Peg Observing Log\label{tab2}}
\tablewidth{0pt}
\tablehead{
\colhead{\bf{Date}} &\colhead{\bf{Exp.time}} &\colhead{\bf{HJD}} &\colhead{\bf{Phase}}\\
\nocolhead{} & \colhead{(s)} & \colhead{(2,450,000+)} & \nocolhead{}
}
\startdata
2016 Jan 22 & 1800 & 7409.944 & 0.259\\
2016 Jan 22 & 1800 & 7409.968 & 0.262\\
2016 Jan 23 & 1800 & 7410.943 & 0.407\\
2016 Jan 23 & 1800 & 7410.966 & 0.410\\
2016 Jan 24 & 1800 & 7411.949 & 0.557\\
2016 Jan 24 & 1800 & 7411.972 & 0.560\\
2016 Jan 25 & 1800 & 7412.935 & 0.703\\
2016 Jan 25 & 1200 & 7412.992 & 0.712\\
2016 Jan 27 & 1200 & 7414.931 & 0.000\\
2016 Jan 27 & 1200 & 7415.016 & 0.013\\
2016 Jan 28 & 1800 & 7415.937 & 0.150\\
2016 Jan 28 & 1800 & 7416.012 & 0.161\\
2016 Jan 29 & 1800 & 7416.939 & 0.299\\
2016 Jan 29 & 1800 & 7416.962 & 0.302\\
2016 Jan 30 & 1800 & 7417.971 & 0.452\\
2016 Jan 30 & 1800 & 7417.995 & 0.456\\
2016 Jan 31  & 1800 & 7418.941 & 0.597\\
2016 Jan 31  & 1200 & 7418.985 & 0.603\\
2016 Jan 31  & 1800 & 7419.012 & 0.607\\
\enddata
\tablecomments{HJD and phase are calculated for mid-exposure of each observation.}
\end{deluxetable}

\section{Spectral analysis}\label{sec3}
Several chromospheric activity indicators, namely $\mbox{Ca~{\sc ii}}$ IRT, $\mbox{He~{\sc i}}$~$\lambda$6678, H$_{\alpha}$, $\mbox{Na~{\sc i}}$ D$_{1}$, D$_{2}$ doublet, $\mbox{He~{\sc i}}$ D$_{3}$, and H$_{\beta}$ lines, formed in a wide range of atmospheric heights from the region of temperature minimum to the upper chromosphere, are covered in our echelle spectra and therefore analyzed, especially the H$_{\alpha}$ line.

As shown in Figure~\ref{Fig1}, clear central emission features appear in the cores of the Ca II IRT absorption line profiles. Moreover, the H$_{\alpha}$ line is always in emission above the continuum and has a significant variable profile during our observation. The $\mbox{Na~{\sc i}}$ D$_{1}$, D$_{2}$ doublet lines are characterized by deep absorption features. The H$_{\beta}$ line often shows strong filled-in features and exhibits emission above the continuum for some of our spectra when the $\mbox{He~{\sc i}}$ D$_{3}$ line shows strong emission.

\subsection{Synthesized spectral subtraction}
\begin{figure}
\includegraphics[width=8.5cm,height=14cm]{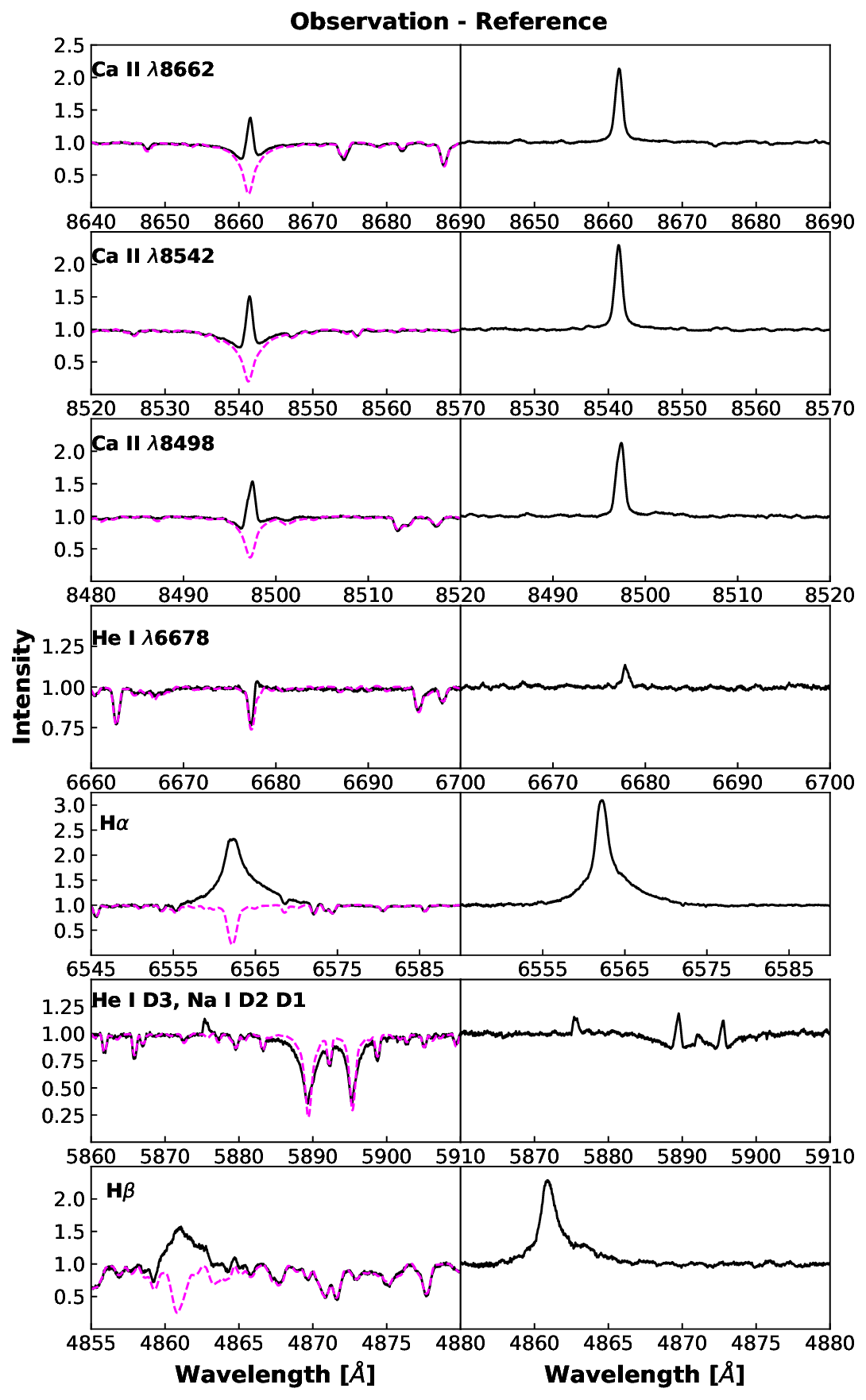}
\caption{Examples of the observed, synthesized, and subtracted spectra for the $\mbox{Ca~{\sc ii}}$~$\lambda$8662, $\mbox{Ca~{\sc ii}}$~$\lambda$8542, and $\mbox{Ca~{\sc ii}}$~$\lambda$8498, $\mbox{He~{\sc i}}$~$\lambda$6678, H$_{\alpha}$, $\mbox{Na~{\sc i}}$ D$_{1}$, D$_{2}$ doublet, $\mbox{He~{\sc i}}$ D$_{3}$, and H$_{\beta}$ line spectral regions, shown for one spectrum obtained at phase 0.597 on 2016 January~31. For the left part of each panel, the black solid line is the observed spectrum, and the magenta dashed line represents the synthesized spectrum constructed from spectrum of a reference star. The resulting subtraction spectrum is displayed in the right part of each panel. The label identifying each chromospheric activity indicator is marked in the corresponding panel.}
\label{Fig1}
\end{figure}
\begin{figure*}
\includegraphics[width=18cm,height=14cm]{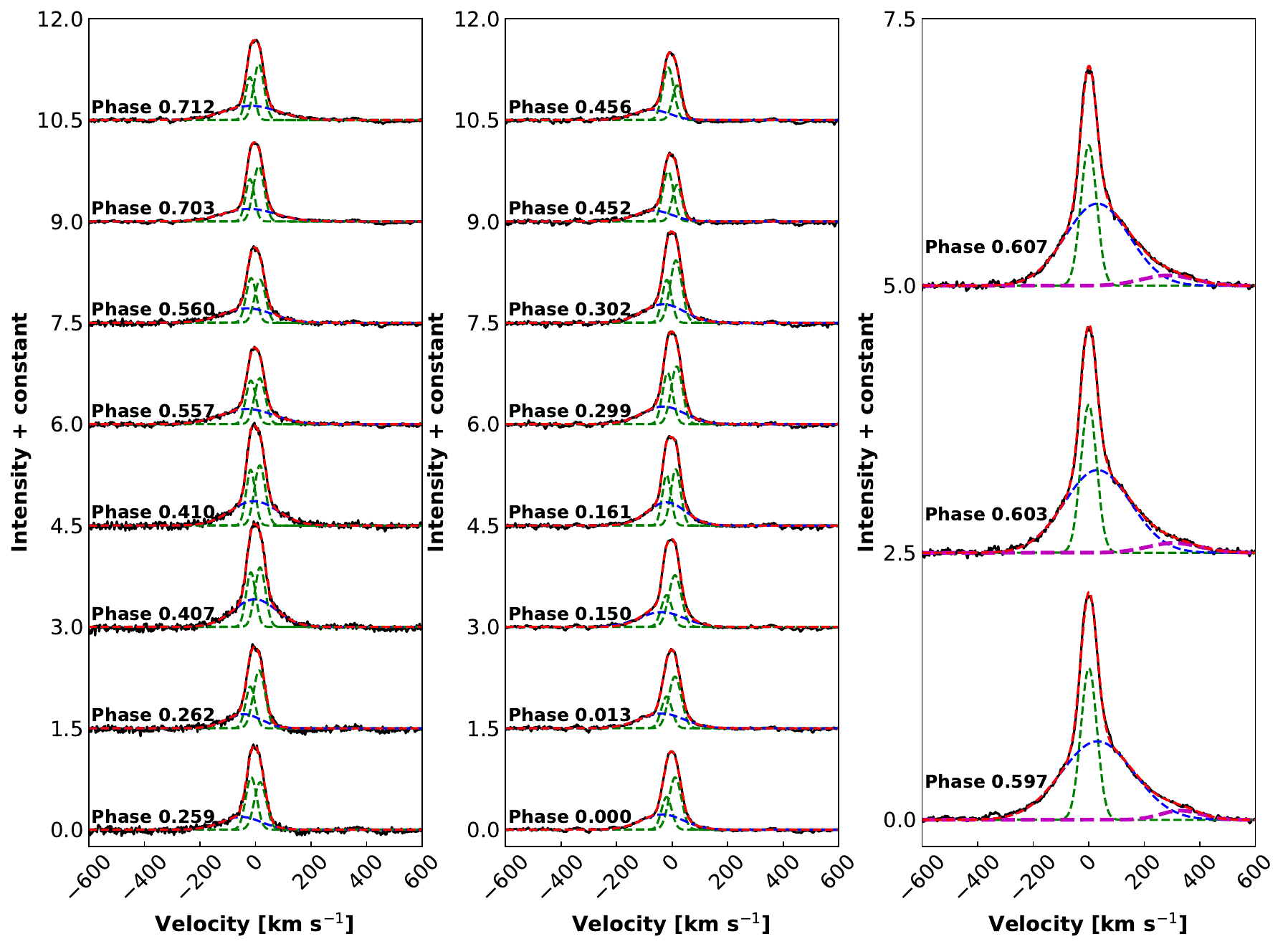}
\caption{Gaussian fitting (dashed lines) for the subtracted H$_{\alpha}$ line profiles (black solid lines). The corresponding phases are marked, and all the spectra are corrected to the rest velocity frame of the K2~IV primary component of II~Peg.}
\label{Fig2}
\end{figure*}
\begin{figure*}
\includegraphics[width=9.cm,height=13cm]{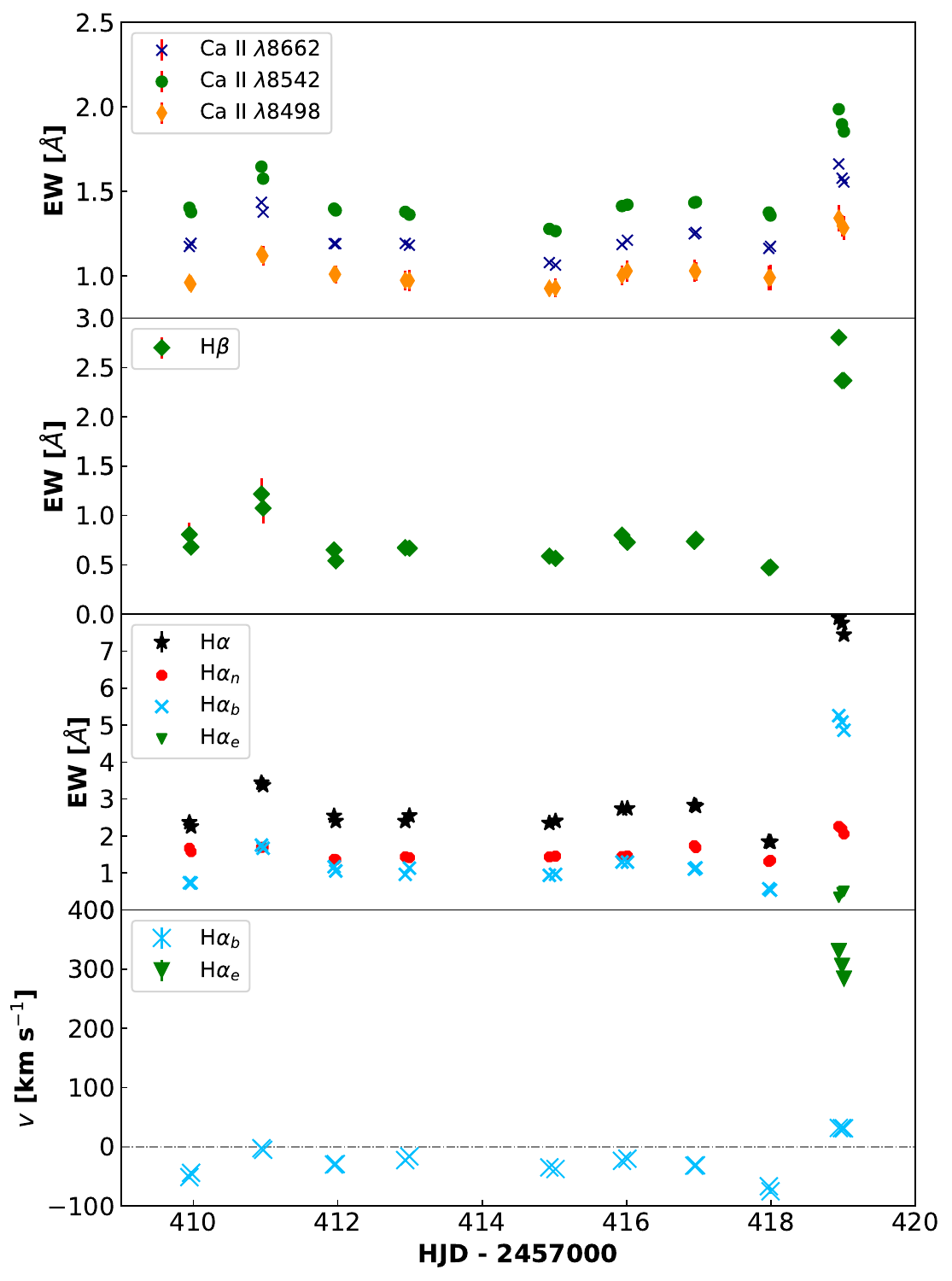}
\includegraphics[width=9.cm,height=13cm]{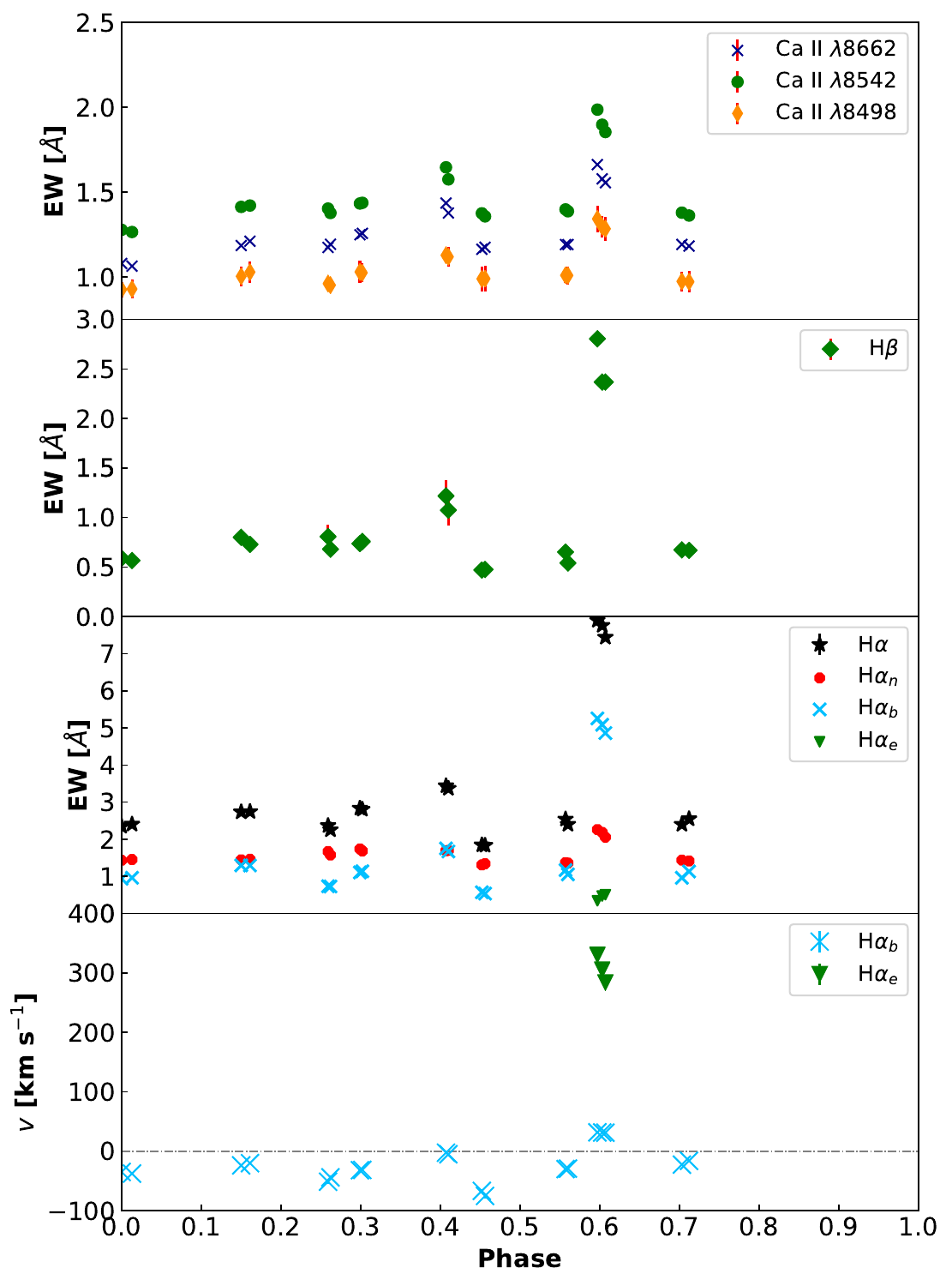}
\caption{Measurements of the subtracted $\mbox{Ca~{\sc ii}}$ IRT, H$_{\alpha}$, and H$_{\beta}$ line profiles vs. HJD (left panel) and orbital phase (right panel). For each panel, from top to bottom: the EWs of the subtracted $\mbox{Ca~{\sc ii}}$ IRT line profiles, the EWs of the subtracted H$_{\beta}$ line profiles, the EWs of the subtracted total H$_{\alpha}$ line profiles and the fitted narrow, broad, and extra emission components, and the velocities of the broad and extra emission components in the subtracted H$_{\alpha}$ line profiles. H$_{\alpha}$$_{e}$ represents the extra emission component in the subtracted H$_{\alpha}$ line profile. The gray dashed-dotted line indicates the zero-point of velocity.}
\label{Fig3}
\end{figure*}
To determine the pure activity contribution in these chromospheric sensitivity lines, we apply a spectral subtraction technique to remove the underlying photospheric contribution by making use of the STARMOD program \citep{Barden1985, Montes1997, Montes2000}. We have widely used this technique for chromospheric activity studies and the detection of prominence-like events in active binary systems \citep[e.g.,][]{Gu2002, Cao2015, Cao2019, Cao2022}.

Since II~Peg is a single-lined binary star, we use only one reference star spectrum to reproduce the synthesized spectrum. By comparison, the spectrum of the inactive star HR~3351 (K0~IV) is used as reference in the synthesized spectrum construction. The rotational velocity ($vsini$) value is determined by using the reference spectrum based on the method described in detail by \citet{Barden1985}. The value of 21.6~km~s$^{-1}$ is obtained from high signal to noise ratio (S/N) spectra, spanning the wavelength regions 6330--6500~\AA~with many photospheric absorption lines, which is in good agreement with the results estimated by several authors \citep[e.g.,][]{Berdyugina1998, Rosen2015, Strassmeier2019}. Consequently, the synthesized spectra are constructed by rotationally broadening the reference spectrum to the $vsini$ value derived above and shifting along the radial-velocity axis, to get best fits. Then, the subtracted spectra between the observed and synthesized ones are calculated, which represent the pure activity contribution of II~Peg. Here, we can also obtain the radial velocities of the K2~IV primary star relative to the reference star at each observation phase, and these values are then used to calibrate the radial velocity variations due to orbital motion in the following analysis.

As a typical example, we illustrate the above-described processing in Fig.~\ref{Fig1}, too. The synthesized spectra match the observational spectra quite well except the $\mbox{Na~{\sc i}}$ D$_{1}$, D$_{2}$ doublet lines. Because they are sensitive to the effective temperature, a slight temperature difference between HR~3351 and II~Peg would produce larger changes in the wings of the $\mbox{Na~{\sc i}}$ D$_{1}$, D$_{2}$ doublet line profiles.

The equivalent widths (EWs) of the subtracted $\mbox{Ca~{\sc ii}}$ IRT, H$_{\alpha}$, and H$_{\beta}$ line profiles are measured, as described in our previous papers \citep{Cao2015, Cao2017, Cao2019}, and summarized in Table~\ref{taba1} together with their uncertainties. The ratios of the EW($\lambda$8542)/EW($\lambda$8498) and the E$_{H{\alpha}}$/E$_{H{\beta}}$ are also provided in Table~\ref{taba1}. The E$_{H{\alpha}}$/E$_{H{\beta}}$ ratios are calculated from the EW(H$_{\alpha}$)/EW(H$_{\beta}$) values with the correction \citep[see][]{Hall1992}, which takes into account the absolute flux density in H$_{\alpha}$ and H$_{\beta}$ lines, and the color index.

During our observation of II~Peg, the EW$_{8542}$/EW$_{8498}$ ratios are in the range of 1--2, which suggests that the $\mbox{Ca~{\sc ii}}$ IRT line emission arises predominantly from plage-like regions \citep[e.g.,][]{Montes2000, Gu2002, Cao2015, Cao2023}. The E$_{H{\alpha}}$/E$_{H{\beta}}$ ratios are usually used as diagnostic for discriminating between prominences and plages on the stellar surface. According to the results of \citet{Hall1992}, the low ratios ($\sim$ 1--2) could be achieved both in plages and prominences viewed against the stellar disk, but high values ($\sim$ 3--15) could only be achieved in prominence-like structures viewed off the stellar limb. Therefore, the high ratio ($\ga$~3) that we found in II~Peg suggests that the emission probably arose from extended prominence-like regions.

With the H$_{\alpha}$ subtraction, showing in Fig.~\ref{Fig2}, we model the line profile by Gaussian fitting. In general, the flaring H$_{\alpha}$ spectra could not simply be fitted by Gaussian profile, as their wing can be well fitted with Lorenzian-like profile due to the strong Stark broadening \citep{Kowalski2017, Namekata2020}. For our situation, it can be seen that the profile may be well fitted by using several Gaussian profile components. In addition to the narrow components, there is a blue- or redshifted broad component having a full width at half maximum (FWHM) of 150--307~km~s$^{-1}$. For most of the subtracted spectra, two narrow components are needed in fitting. This is reasonable, because the distribution of chromospheric activity regions is not uniformly expressed on the stellar surface. Moreover, for spectra taken at phases 0.597, 0.603, and 0.607 on 2016~January~31 (see the rightmost panel of Fig.~\ref{Fig2}), it is noteworthy that there is a far-redshifted extra emission component (shown by the magenta dashed-dotted line) in the subtraction, except a narrow and a strong redshifted broad components. Here, just one narrow component is used in fitting, probably because a large flare region dominates the activity of II~Peg at these observing phases. In Table~\ref{taba1}, we also list the EWs of the broad and narrow components, and the velocities of the broad component.

Finally, we plot the measurements listed in Table~\ref{taba1} as functions of HJD and orbital phase in Fig.~\ref{Fig3}, respectively, together with the EWs, and velocities of the far-redshifted extra emission component in the subtracted H$_{\alpha}$ line. It can be seen that the EW variations of the different chromospheric activity indicators are closely correlated, similar to the behavior of many chromospherically active stars.

\subsection{Mean spectrum subtraction}
\begin{figure}
\includegraphics[width=8.5cm,height=14cm]{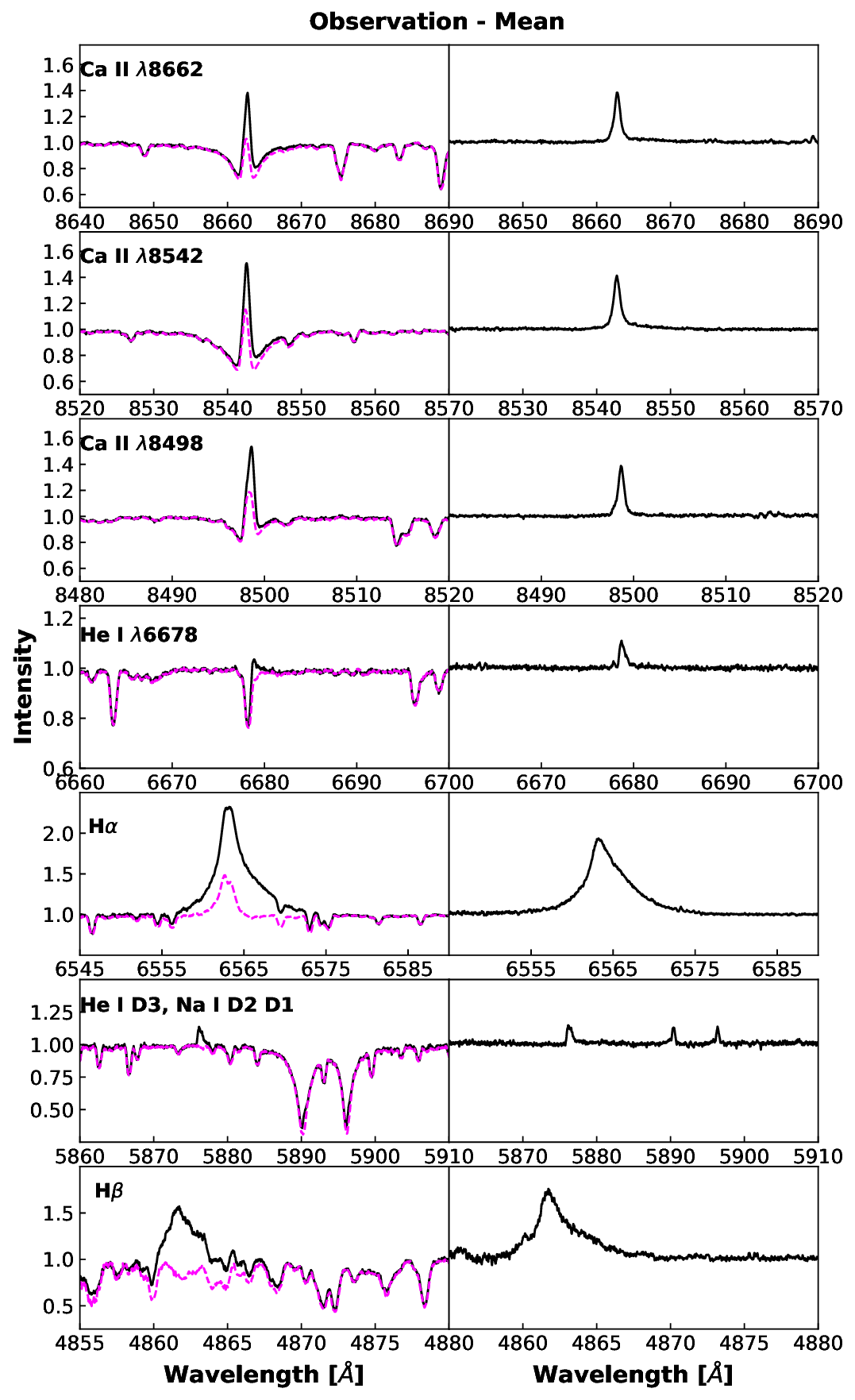}
\caption{Same as Fig.~\ref{Fig1}, but for subtracting a mean spectrum (magenta dashed line) from the observed spectrum for each chromospheric activity indicator.}
\label{Fig4}
\end{figure}
\begin{figure*}
\includegraphics[width=18cm,height=12cm]{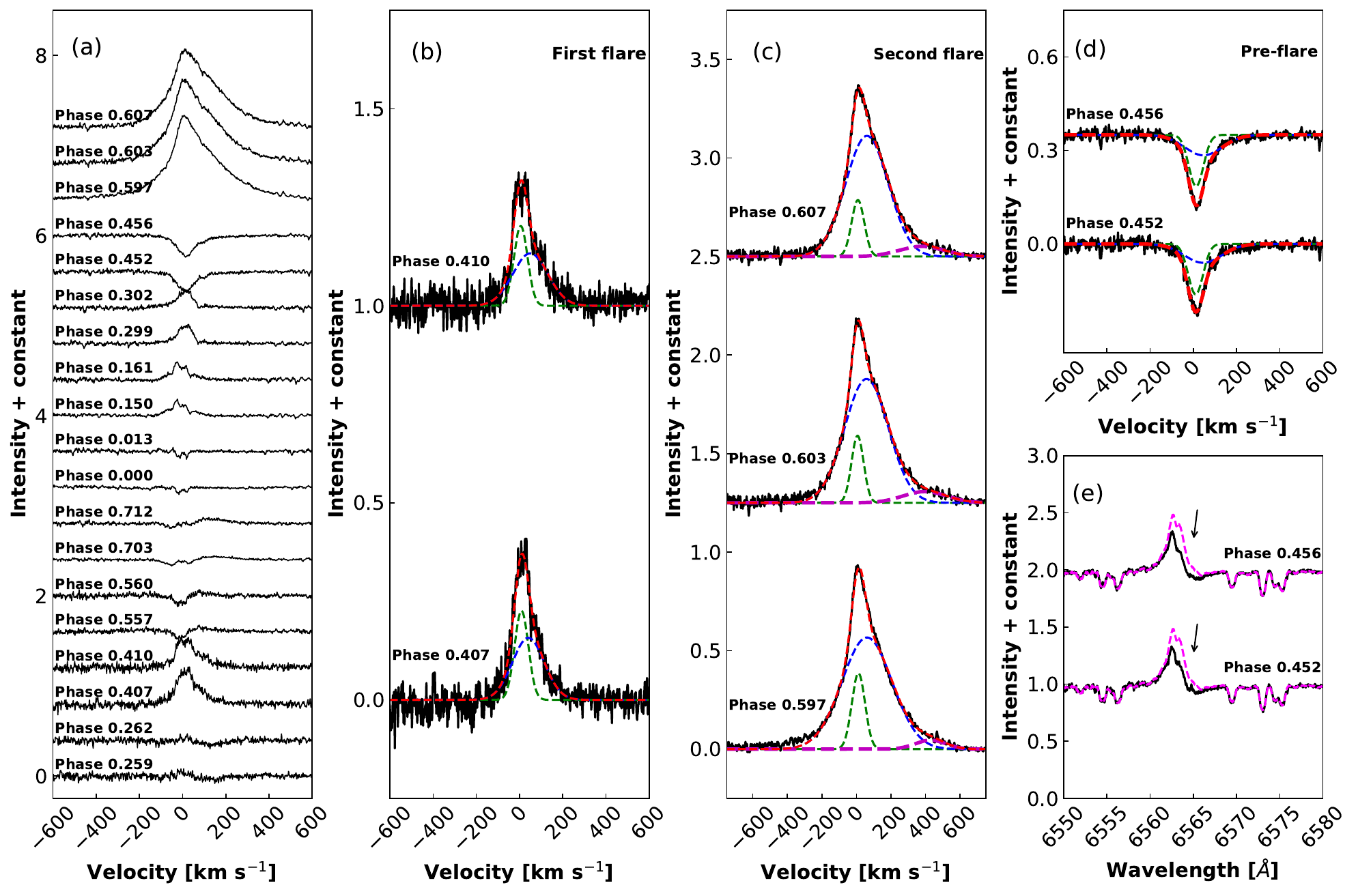}
\caption{The H$_{\alpha}$ residuals between the observed and mean spectra. (a) All residuals at different phases during our observation, (b) and (c) Gaussian fitting (dashed lines) for the residuals during two flares (solid lines), (d) Gaussian fitting (dashed lines) for the residuals during pre-flare, (e) the observed pre-flare spectra (solid lines) compared with the mean spectrum (magenta dashed lines). The corresponding phases are marked in each panel, and arrows indicate extra absorption features in the red wings of the observed H$_{\alpha}$ line profiles. The residual spectra are corrected to the rest velocity frame of the K2~IV primary component of II~Peg.}
\label{Fig5}
\end{figure*}
There are two strong optical flare events detected during our observation, which are discussed in Sec.~4.1 in more detail. Typically, a pre-flare spectrum subtraction is performed to analyze the flaring spectra \citep[e.g.,][]{Namekata2022, Inoue2023}. In our situation, however, because the closest pre-flaring spectra were taken the night before, and because the chromospheric line profile of II Peg would change significantly between different nights, we subtract a mean spectrum from the observed spectra to analyze the behavior of the chromospheric lines during the flares and to identify whether the H$_{\alpha}$ asymmetry is associated with the flares. We first correct all the spectra so that they are aligned, and then the mean spectrum is obtained from spectra except ones taken during the two flares. Therefore, the mean spectrum could represent the mean activity state of II~Peg when no flare happens. To a certain extent, the residuals of the two flares can reflect the flare behavior also the flare-related activity phenomena. An example for illustrating this process is displayed in Fig.~\ref{Fig4}.

Fig.~\ref{Fig5}(a) exhibits all the H$_{\alpha}$ residuals between the observed and mean spectra, and as shown in Fig.~\ref{Fig5}(b)--(d), we model the residual profiles obtained during two flares (observed on 2016 January~23 and 31, respectively) and the pre-flare of the second flare (observed on 2016 January~30) by using Gaussian fitting. For the first flare, the residual spectra are modeled by a narrow Gaussian profile component and a redshifted broad component with a bulk velocity of about 40--50~km~s$^{-1}$. The pre-flare residuals of the second flare show absorption features and red asymmetry, which could be modeled by a narrow Gaussian absorption component and a slightly redshifted broad absorption component. The absorption feature is not just because II~Peg has a lower chromospheric activity at these observing phases when comparing with the mean activity state. A probable reason is that there is an extra absorption in the red wing of the H$_{\alpha}$ line profile, as indicated through comparing the observed and mean spectra in Fig.~\ref{Fig5}(e). For the second flare, the residual spectra are similar with the STARMOD subtraction, which still can be modeled by using a weak narrow, a redshifted strong broad, and a far-redshifted extra emission components. The bulk velocity of the redshifted broad emission component is around of 60~km~s$^{-1}$, while the far-redshifted extra emission component has a bulk velocity decreased from 429~km~s$^{-1}$ to 376~km~s$^{-1}$, and could extends its velocity up to 600--650~km~s$^{-1}$ during our observations. Moreover, the extra emission becomes more and more prominent, with the FWHM increased from 206~km~s$^{-1}$ to 307~km~s$^{-1}$.

On the whole, during the two flares, our analytical method demonstrates that the H$_{\alpha}$ red asymmetry is closely related to flaring activity. Hereafter, all investigations and calculations for the flare and flare-associated (especially the possible CME event discussed in Section.~4.2) events are based on the results of mean spectrum subtraction.

\section{Results and discussion}\label{sec4}
\subsection{Flares and flare energies}
\begin{deluxetable}{lccc}
\tablenum{3}
\tablecaption{EWs, Surface Fluxes, and Flare Luminosities of the Residuals of Chromospheric Activity Lines at Flare Maximums\label{tab3}}
\tablewidth{0pt}
\tablehead{
\colhead{\bf{Line}}&\colhead{\bf{EW}}&\colhead{\bf{Fs ($\times~10^{7}$)}}&\colhead{\bf{L ($\times~10^{30}$)}}\\
\nocolhead{}&\colhead{({\AA})}&\colhead{(erg~cm$^{-2}$~s$^{-1}$)}&\colhead{(erg~s$^{-1}$)} 
}
\startdata
\multicolumn4c{\bf{Flare on 2016~Jan~23}}\\
$\mbox{Ca~{\sc ii}}~\lambda8662$ & 0.208$\pm$0.015 & 0.04 & 0.30 \\
$\mbox{Ca~{\sc ii}}~\lambda8542$ & 0.215$\pm$0.014 & 0.05 & 0.32\\
$\mbox{Ca~{\sc ii}}~\lambda8498$ & 0.114$\pm$0.033 & 0.02 & 0.17\\
$\mbox{He~{\sc i}}~\lambda6678$  & 0.047$\pm$0.012 & 0.01 & 0.09 \\
H$_{\alpha}$                                   & 0.887$\pm$0.097 & 0.23 & 1.65\\
H$_{\beta}$                                     & 0.705$\pm$0.082 & 0.17 & 1.16\\
&&&\\
\bf{Total lines}  &  &    & 3.69 \\
\hline
\multicolumn4c{\bf{Flare on 2016~Jan~31}}\\
$\mbox{Ca~{\sc ii}}~\lambda8662$ & 0.513$\pm$0.002 & 0.11 & 0.74\\
$\mbox{Ca~{\sc ii}}~\lambda8542$ & 0.622$\pm$0.002 & 0.13 & 0.92\\
$\mbox{Ca~{\sc ii}}~\lambda8498$ & 0.394$\pm$0.004 & 0.08 & 0.59\\
$\mbox{He~{\sc i}}~\lambda6678$  & 0.093$\pm$0.014 & 0.02 & 0.17\\
H$_{\alpha}$                                   & 5.714$\pm$0.145 & 1.51 & 10.63\\
$\mbox{Na~{\sc i}}$~D$_{1}$         & 0.075$\pm$0.019 & 0.02 & 0.14\\
$\mbox{Na~{\sc i}}$~D$_{2}$         & 0.086$\pm$0.023 & 0.02 & 0.16\\
$\mbox{He~{\sc i}}$~D$_{3}$         & 0.155$\pm$0.015 & 0.04 & 0.29 \\
H$_{\beta}$                                     & 2.467$\pm$0.033 & 0.58 & 4.06\\
&&&\\
\bf{Total lines}  &  &  & 17.11 \\
\enddata
\end{deluxetable}
Stellar flares are dramatic explosive phenomena in the outer atmosphere of active stars, commonly believed to be caused by the energy released in magnetic reconnection. Flares can be observed across the entire electromagnetic spectrum from shorter X-ray to longer radio wavelengths. In the optical spectral region, the $\mbox{He~{\sc i}}$ D$_{3}$ line is an important and commonly used indicator to trace flare activity in solar and stellar chromospheres because of its very high excitation potential. When flares happen, the $\mbox{He~{\sc i}}$ D$_{3}$ line shows an obvious emission feature above the continuum level, as widely observed during flares on the Sun \citep{Zirin1988} and several RS CVn-type stars \citep[e.g.,][]{Gu2002, Garcia2003, Cao2019}.

Two optical flare events were detected during our observations. The first flare was observed at phases 0.407 and 0.410 on 2016~January~23, while the second one occurred at phases 0.597, 0.603, and 0.607 on 2016~January~31. The second flare is much stronger than the first one (see Table~\ref{tab3} and Figure~\ref{Fig3}). During two optical flares, the $\mbox{He~{\sc i}}$ D$_{3}$ line shows an obvious emission feature and the H$_{\alpha}$ line and other activity lines display stronger emission. Moreover, another $\mbox{He~{\sc i}}$ line, $\lambda$6678, also shows a distinct emission feature, especially after subtracting the synthesized spectrum or the mean spectrum. A typical example of flaring spectrum could be found in Figure~\ref{Fig1} or Figure~\ref{Fig4}. 

We did not capture the whole life cycle of the two flares of II~Peg due to limited observations. Since the flare intensity shows a decreasing trend, our observations were part of the gradual decay phases of flares and the observations at phases 0.407 and 0.597 could be considered as the emission maxima of two flares. We compute the stellar continuum flux $F_{H{\alpha}}$~(in erg cm$^{-2}$ s$^{-1}$ \AA$^{-1}$) near the H$_{\alpha}$ line as a function of the color index $B-V$ ($\sim$~1.031 for II~Peg; \citealt{Messina2008}) based on the empirical relationship: 
\begin{eqnarray}
\log{F_{H{\alpha}}}=[7.538-1.081(B-V)]\pm{0.33}\nonumber \\
0.0~\leq~B-V~\leq~1.4
\end{eqnarray}
of \citet{Hall1996}, and then convert the EW into an absolute surface flux $F_{S}$~(in erg~cm$^{-2}$~s$^{-1}$). Following the method adopted by \citet{Montes1999}, \citet{Garcia2003}, and \citet{Cao2019}, we obtain the stellar continuum fluxes $F_{\lambda}$ for the other chromospheric activity lines by using the flux ratios: 
\begin{eqnarray}
\frac{F_{6563}}{F_{\lambda}}~=~\frac{B_{6563}(T)}{B_{\lambda}(T)}
\end{eqnarray}
where B($\lambda$, T) is the Plank function. The flux ratios are given by assuming a blackbody with contribution of the K2~IV primary of II~Peg at the effective temperature $T_{eff}$~=~4750~K. Then, we convert these fluxes into luminosities using the stellar radius $R_{\star}$~=~3.4$R_{\sun}$. The corresponding EWs, surface fluxes, and luminosities at flare maxima in the residuals of each chromospheric activity line are listed in Table~\ref{tab3}. For the first flare, because the flare emission in the $\mbox{Na~{\sc i}}$ D$_{1}$, D$_{2}$ doublet and $\mbox{He~{\sc i}}$ D$_{3}$ lines is very weak and the spectrum has low S/N, we do not measure them. The H$_{\alpha}$ luminosities of two flares are of similar order of magnitude as strong flares on other highly active RS CVn-type stars, such as V711~Tau \citep{Cao2015}, UX~Ari \citep{Montes1996, Gu2002, Cao2017}, HK~Lac \citep{Catalano1994}, SZ~Psc \citep{Cao2019, Cao2020}, and RS~CVn \citep{Cao2023}, especially for the second flare.

Moreover, in Table~\ref{tab3}, we also provide the total luminosities summed from observed chromospheric activity lines. For our observation, it is impossible to estimate the flare duration from the initial outburst to the end, but we can give a timescale of 1800~seconds for two optical flares, respectively. The duration times of two flares are just the exposure times of our observations, which means they are underestimated. Thus, the estimated total energies released in optical chromospheric lines are $6.6~\times~10^{33}$~erg for the first flare and $3.1~\times~10^{34}$~erg for the second one. These values represent the low limits of the flare energies and are comparable to those of stellar superflares, particularly the second flare.

\subsection{H$_{\alpha}$ profile asymmetry during the flares}
\subsubsection{Interpretation of H$_{\alpha}$ profile asymmetry }
In the case of solar flares, as discussed by \citet{Moschou2019}, \citet{Koller2021}, and \citet{Wu2022}, the redshifted H$_{\alpha}$ emission may arise from coronal rain, or from chromospheric condensation, or from prominence/filament eruption with a backward direction. The red asymmetry of H$_{\alpha}$ profile may be caused by one of these three possible scenarios or their combinations. It is difficult to distinguish them because the events are not spatially resolved. One possible way to identify them is based on the Doppler shifts. Firstly, coronal rain is dense and cool plasma condensation formed in the hot corona, and then falls down along post-flare loops to the solar surface with average velocity of about 60--70~km~s$^{-1}$ \citep{Antolin2012, Lacatus2017}. A flare-driven coronal rain was reported with a velocity up to 134~km~s$^{-1}$ by \citet{Martinez2014} and the downward acceleration is generally not more than 80~m~s$^{-2}$. Secondly, chromospheric condensation is usually observed with a typical downflow velocity of tens of kilometers per second \citep{Ichimoto1984}. And such condensation usually occurs during the impulsive phase of solar flare and is caused by nonthermal heating processes in flares. Finally, the typical velocity of solar prominence/filament eruption is in the range of 10--500~km~s$^{-1}$. And as discussed by \citet{Wu2022}, a detectable backward prominence/filament eruption requires specially spatial location, which implies that its occurrence is very rare.

During our two flares, there are the redshifted broad emission components with a bulk velocity of about 40--50~km~s$^{-1}$ for the first flare and about 60~km~s$^{-1}$ for the second one. The velocities of the redshifted broad components correspond well  to the velocity of chromospheric condensation and coronal rain in the solar case. This suggests that chromospheric condensation or coronal rain is responsible for their occurrence.

During the second flare, the far-redshifted extra emission component has a large velocity decreased from 429~km~s$^{-1}$ to 376~km~s$^{-1}$ and could extends its velocity up to 600--650~km~s$^{-1}$, which is significant larger than the typical velocities of coronal rain and chromospheric condensation in solar case. And, the velocity approaches the larger value of solar prominence/filament eruption. Thus, stellar prominence eruption might be a good explanation for its origination based on what is known about the Sun. Moreover, there is a scenario for supporting its occurrence. As analyzed in Section~3.2, for the pre-flare spectra, absorption features appear in the residuals. Part of the absorption, especially the extra absorption in the red wing of the H$_{\alpha}$ line, could be caused by a stellar prominence projected on the right hemisphere of the star, and therefore scattered the underlying chromospheric activity emission out of the line of sight \citep{Collier1989}. On the next observing night, after the K2~IV primary star rotated about 50$^{\circ}$ from phase 0.456 to phase 0.597, it is possible that the corotated prominence moved close to the limb of the stellar disk and then occurred a backward eruption due to a superflare. An alternate scenario can be that the prominence just moved to the opposite hemisphere, and then a forward eruption occurred. When the eruption began to extend outward, we just observed them. In both cases, the flare must happened around the limb of the stellar disk, where at least one of its footpoints which produced H$_{\alpha}$ emission can be seen by us. Just at these specially spatial locations, stellar prominence eruption could result in a redshifted emission feature, which is consistent with the situation discussed by \citet{Wu2022} and the scenario depicted in the second panel of Figure~4 of \citet{Moschou2019}, respectively.

Prominence/filament eruption can lead to CME, especially when the eruption velocity is sufficiently large. In a superflare, for example, \cite{Inoue2023} detected a blueshifted excess emission component of the H$_{\alpha}$ line on RS~CVn-type star V1355~Orionis. Because the velocity greatly exceeds the star's escape velocity, it could be thought to originate from prominence eruption and could develop into a CME. Due to projection effects, however, our measured velocity only is the lower limit of the true velocity of eruption. For the eruption scenarios discussed above, there may be a larger angle between the eruption direction and the line of sight, and therefore the prominence eruption has a small velocity component along the line of sight. For the K2~IV primary star of II~Peg, the escape velocity $v_{esp}~=~630(\frac{M_{\star}}{M_{\sun}})^{1/2}(\frac{R_{\star}}{R_{\sun}})^{-1/2}$~km~s$^{-1}$ can be determined to be about 306~km~s$^{-1}$ at the surface of the star. Therefore, for our situation, the underestimated velocity is still much larger than the star's escape velocity, which provides a very important evidence that this stellar prominence eruption can develop into a successful CME event.

The energy scale of the flares on II~Peg vastly exceeds that of typical solar flares. This suggests that, while similar mechanisms may operate during the flares, the magnitude of redshift phenomena could differ greatly between solar flares and flares on stars. \citet{Namizaki2023ApJ.945.61N} observed the H$_{\alpha}$ red asymmetry during a superflare on  an active M dwarf, YZ Canis Minoris. The velocity of the red asymmetry is much fast ranging from 200--500~km~s$^{-1}$, which could be interpreted as chromospheric condensation heated by nonthermal electrons. Through the numerical simulations, moreover, \citet{Longcope2014} suggested that strong heating fluxes during large superflares could produce fast chromospheric condensation. For our situation, the far-redshifted extra emission is accompanied by the prominent line broadening (see Section~3.2), which probably suggests the appearance of chromospheric condensation due to nonthermal heating during the flare decay. Although the velocity of the far-redshifted extra emission is much high, therefore, we can not entirely rule out the possibility of chromospheric condensation.

In addition, it is estimated that the flare loop size is considerably larger in the case of superflares on RS CVn-type stars \citep[e.g,][]{Tsuboi2016}. Consequently, the velocity of falling plasma along loops during coronal rain for RS CVn-type stars could be much higher than those in solar flares. In the solar case, the falling materials move downward along loops with an acceleration of about $g_{eff}$/3 and its maximum velocity is about 100~km~s$^{-1}$ \citep{Antolin2020}. According to the method of \citet{Wu2022}, the maximum velocity of the falling plasma on II~Peg is estimated to be about 63~km~s$^{-1}$ on the assumption that the propagation distance scales with the radius of the star. \citet{Tsuboi2016} proposed that the largest loop size from II~Peg are much larger than the binary separation. If we assume that the propagation distance of coronal rain is the binary separation of II~Peg ($\sim$15.6~R$_{\sun}$), the maximum velocity is about 134~km~s$^{-1}$. The velocity of falling plasma is much smaller than the velocity of the far-redshifted emission during the flare of II~Peg, and therefore we could eliminate the possibility of coronal rain.

\subsubsection{Mass of the possible CME}
By following \citet{Houdebine1990} and \citet{Koller2021}, mass of the CME can be estimated as:
\begin{eqnarray}
M_{CME}~\geqslant~\frac{4{\pi}d^{2}f_{line}m_{H}}{A_{ji}h\nu_{ji}P_{esc}}\frac{N_{tot}}{N_{j}}
\end{eqnarray}
in which $d$ is the distance from the star, $f_{line}$ is the corresponding line flux from the CME emission feature, $m_{H}$ is the mass of a hydrogen atom, $N_{tot}/N_{j}$ is the ratio between the number density of hydrogen atoms and the number density of hydrogen atoms at excited level $j$, $h$ is the Planck constant, $\nu_{ji}$ and $A_{ji}$ are the frequency and Einstein coefficient for a spontaneous decay from level $j$ to $i$, and $P_{esc}$ is the escape probability.

In the calculation, we use the same parameter settings and same strategy as \citet{Koller2021}. Due to a lack of $N_{tot} /N_{j}$ value for the H$_{\alpha}$ line ($j$~= 3, $i$ =~2) in the literature, they estimated $M_{CME}$ from the H$_{\gamma}$ line ($j$~= 5, $i$ =~2) flux and then transformed to the H$_{\alpha}$ line by adopting a Balmer decrement of three \citep{Butler1988}. \citet{Houdebine1990} performed non-LTE modeling for a well-known flaring dMe star AD~Leo and estimated $N_{tot}/N_{5}$ = $2~\times~10^{9}$. A typical value of 0.5 is used for $P_{esc}$, which indicates that half of the radiation escapes from the moving feature \citep{Leitzinger2014}. $A_{ji}$ is $2.53~\times~10^{6}$ for the H$_{\gamma}$ line \citep{Wiese2009}.

Adopting the same strategy as \citet{Wu2022}, we use the luminosity $L_{line}$ to instead of $4{\pi}d^{2}f_{line}$, because we also do not have absolute flux calibration. We have obtained the EW and luminosity of H$_{\alpha}$ at flare maximum (see Table~\ref{tab3}), and then derived the $L_{line}$ according to the EW of the far red-shifted extra emission component. Finally, we calculate the mass of CME, which is in the range of 0.83--1.48~$\times~10^{20}$~g from our three exposures.

\subsubsection{Comparison with other events}
\begin{figure*}
\includegraphics[width=9.cm,height=7.75cm]{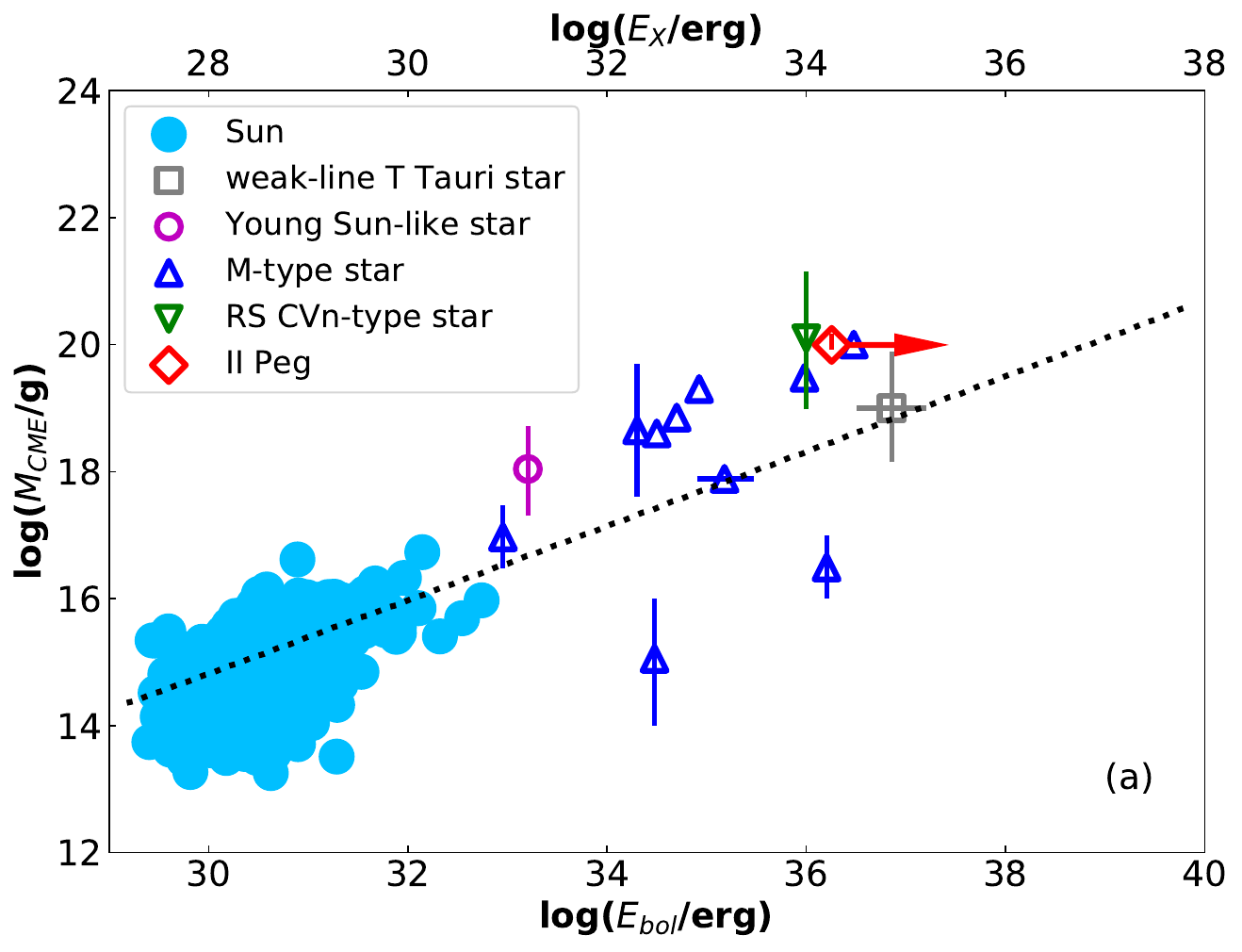}
\includegraphics[width=9.cm,height=7.75cm]{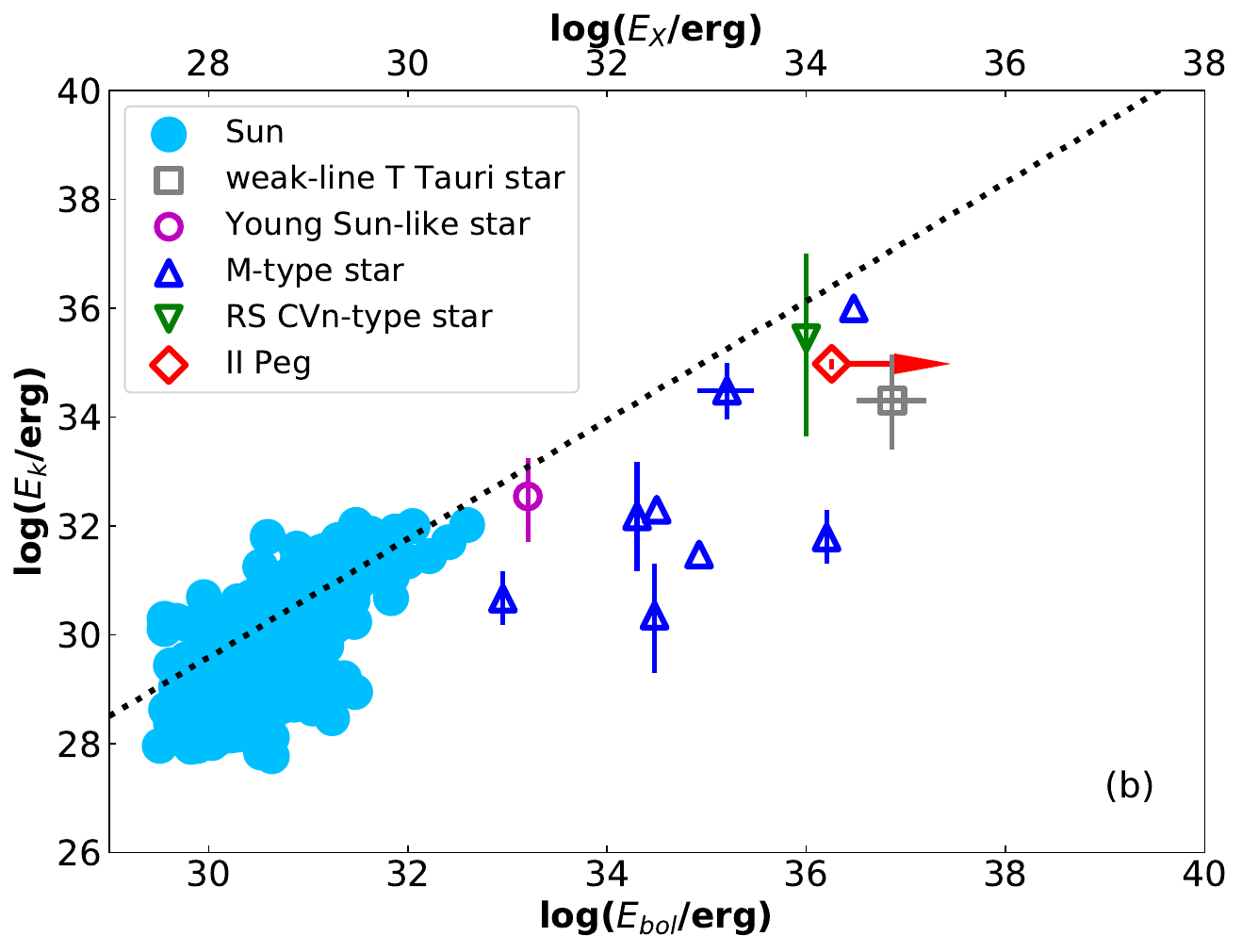}
\caption{(a) CME mass $M_{CME}$ and (b) kinetic energy $E_{k}$ are plotted as functions of flaring bolometric energy (bottom horizontal axis) and X-ray energy emitted in the GOES band (top axis). The flare energy of II~Peg is the lower limit, which therefore is indicated by an arrow in the direction of getting larger. The Solar events are derived from \citet{Yashiro2009} and indicated by filled sky-blue circles. The trend fits to the solar data obtained in \citet{Drake2013} are presented by the dotted lines, which could be expressed as  $M_{CME}$~$\propto$~$E^{0.59}$ and $E_{k}$~$\propto$~$E^{1.05}$, respectively. Stellar CME candidates on different type stars are obtained from \citet{Moschou2019}, \citet{Wang2021, Wang2022}, \citet{Namekata2022}, and \citet{Inoue2023}.}
\label{Fig6}
\end{figure*}
To compare the detected CME with other events, we convert the flare energy emitted in the H$_{\alpha}$ line to the GOES X-ray (1--8~\AA~band) using the scaling relation between H$_{\alpha}$ and GOES soft X-ray flare energies in Figure~2 and Equation~(1) of \citet{Haisch1989}, and then use a correction of $E_{bol}$ = 100$E_{X}$ to obtain the bolometric flare energy. Therefore, the lower limit of the bolometric flare energy, $E_{bol}(H_{\alpha})$, is estimated to be 1.8~$\times~10^{36}$~erg. Moreover, with the estimated mass and the corresponding velocity, the kinetic energy of CME is inferred to be about 0.76--1.15~$\times~10^{35}$~erg during our observation.

We plot the CME mass and kinetic energy as functions of flare energy emitted in the GOES X-ray and bolometric flare energy in Figure~\ref{Fig6}, which also includes (1) solar CME events studied in \citet{Yashiro2009}, (2) stellar CME candidates detected by using Doppler shift method and gathered by \citet {Moschou2019} in the literature, (3) possible flare-associated CME candidates on M-dwarf stars \citep{Wang2021, Wang2022}, (4) an eruptive filament event on a young solar-type star EK~Dra \citep{Namekata2022}, (5) a high-velocity prominence eruption leading to a flare-associated CME on the RS~CVn-type star V1355~Orionis \citep{Inoue2023}. For EK~Dra and V1355~Orionis, the flare energies in GOES X-ray band are also obtained by using the energies emitted in H$_{\alpha}$ and the scaling relation of \citet{Haisch1989} by us.

It can be seen from the figure that our estimated mass is slightly larger than the value predicted from the solar flare-CME extrapolation, which is similar to the result of RS~CVn-type star V1355~Orionis. Taking into account the fact that the flare energy of II~Peg is the low limit, we can not say that our estimated mass is inconsistent with the scaling relation of solar events. Moreover, similar to several CME candidates on other active stars, the obtained CME kinetic energy is less than the one extrapolated from solar events. This mainly because the prominence eruptions in stars generally have a much smaller velocity than CMEs on the Sun \citep{Maehara2021, Namekata2022}. Furthermore, a suppression mechanism by the the overlying magnetic field in active stars is proposed to explain small kinetic energies \citep[e.g.,][]{Drake2016, Alvarado2018}. Finally, it is possible that the velocity of stellar CME is greatly underestimated, because it is just the projection component along the line of sight.

\subsection{H$_{\alpha}$ profile asymmetry in the cases without the flares}
II~Peg is a very active RS~CVn-type star with variable H$_{\alpha}$ emission above the continuum. After using the synthesized spectral subtraction technique, the subtracted H$_{\alpha}$ profile presents broad wings. When no strong flares happened, it is worth to notice that the broad emission component is always blueshifted (see Table~\ref{taba1} and Fig~\ref{Fig3}). Broad emission wings had been observed in II~Peg \citep{Montes1997} and in several other chromospheric active stars \citep[e.g.][]{Gu2002, Montes1997, Montes2000, Cao2015}. \citet{Montes1998} analyzed the H$_{\alpha}$ line profile, using the spectral subtraction technique, for chromospherically active binary systems and weak-lined T~Tauri stars. They found that, in some active stars, the subtracted H$_{\alpha}$ line profile has broad wings, which could be interpreted as arising from microflare activity through contrasting with the broad components in the transition region lines.

From Fig~\ref{Fig3}, it can be seen that there is a close correlation between the contribution of the broad component and the level of stellar activity. That is, the EW variation of the subtracted H$_{\alpha}$ line is mainly dominated by the broad emission component. On the other hand, there is also a close correlation between the blueshifted velocity of the broad component and the EW. The more blue the broad emission component shifts to, the weaker the EW becomes.

\section{Summary and conclusions}\label{sec5}
Based on the analysis of high-resolution spectroscopic observation of the very active RS~CVn-type star II~Peg, performed from 2016 January~22 to 31, some interesting results are obtained. Two strong optical flare events were detected on 2016 January~23 and 31, which are demonstrated by the prominent $\mbox{He~{\sc i}}$ D$_{3}$ and $\mbox{He~{\sc i}}$~$\lambda$6678 line emission together with the great enhancements in other chromospheric activity lines. The low limits of the flare energies are estimated to be of the order of $10^{33}$~erg and $10^{34}$~erg, respectively, and have reached the scale of stellar superflare. Based on the results of mean spectrum subtraction during the flares, the H$_{\alpha}$ residual profile shows red asymmetry, whose behavior has a close relation to flaring activity. From what is known about solar flares, chromospheric condensation or coronal rain is a possible origination of the redshifted broad emission components having a velocity of tens of kilometers per second. Most notably, during the second flare, a far-redshifted extra emission component was detected in the H$_{\alpha}$ line profile and can be explained as caused by a possible stellar prominence eruption. Its Doppler velocity greatly exceeds the star's escape velocity, which implies that this prominence eruption can develop into a CME event. The mass and kinetic energy of the CME are estimated to be 0.83--1.48~$\times~10^{20}$~g and 0.76--1.15~$\times~10^{35}$~erg, respectively. The mass is slightly larger than the value expected from the solar flare-CME extrapolation. Similar to several CME candidates on other active stars, the kinetic energy is less than the extrapolated one. In addition, we could not completely rule out the possibility of chromospheric condensation resulting in the far-redshifted extra emission. Finally, when no strong flares happened, there is a blueshifted broad emission component in the STARMOD subtracted H$_{\alpha}$ line profile. Its behavior is closely associated with the H$_{\alpha}$ activity features.

\begin{acknowledgments}
The authors thank the anonymous referee for the careful review and helpful suggestions, which lead to significantly improvement in our manuscript. The authors would like to thank the staff of the Xinglong 2.16 m telescope for their help and support during our observation. This work is partially supported by the Open Project Program of the Key Laboratory of Optical Astronomy, National Astronomical Observatories, Chinese Academy of Sciences. The present study is also financially supported by the National Natural Science Foundation of China (NSFC) under grant Nos. 10373023, 10773027, 11333006, and U1531121, the Yunnan Fundamental Research Projects (grant Nos.~202201AT070186 and 202305AS350009), the Yunnan Revitalization Talent Support Program (Young Talent Project), and International Centre of Supernovae, Yunnan Key Laboratory (No.~202302AN360001). The authors also acknowledge the science research grant from the China Manned Space Project.
\end{acknowledgments}

\appendix
\section{Measurements for the subtracted chromospheric activity indicators}
Table~\ref{taba1} lists the measurements of the subtracted profiles and the fittings of the broad and narrow component of the subtracted H$_{\alpha}$ line profiles.
\begin{longrotatetable}
\begin{deluxetable*}{ccccccccccccc}
\tablenum{A1}
\tablecaption{Measurements of the Subtracted Profiles\label{taba1}}
\tablewidth{0pt}
\tablehead{
\colhead{\bf{HJD}} & \colhead{\bf{Phase}} & \multicolumn7c{\bf{EW({\AA})}} & \colhead{\bf{$v$(H$_{\alpha}$$_{b}$)}} & \colhead{\bf{FWHM(H$_{\alpha}$$_{b}$)}} & \colhead{$\frac{\bf{EW(\lambda8542)}}{\bf{EW(\lambda8498)}}$} & \colhead{$\frac{\bf{E_{H{\alpha}}}}{\bf{E_{H{\beta}}}}$} \\
\cline{3-9}\\
\colhead{(2,450,000+)} & \nocolhead{} & \colhead{$\mbox{Ca~{\sc ii}}$ $\lambda8662$} & \colhead{$\mbox{Ca~{\sc ii}}$ $\lambda8542$} & \colhead{$\mbox{Ca~{\sc ii}}$ $\lambda8498$} & \colhead{H$_{\alpha}$} & \colhead{H$_{\alpha}$$_{n}$} & \colhead{H$_{\alpha}$$_{b}$} & \colhead{H$_{\beta}$} & \colhead{(km~s$^{-1}$)} & \colhead{(km~s$^{-1}$)} & \nocolhead{} & \nocolhead{}\\
}
\startdata
7409.944&0.259&1.174$\pm$0.001&1.403$\pm$0.024&0.961$\pm$0.036&2.368$\pm$0.077&1.671&0.736&0.807$\pm$0.122&-51.33$\pm$3.96&167.29$\pm$5.70&1.46&4.13\\
7409.968&0.262&1.193$\pm$0.003&1.377$\pm$0.022&0.952$\pm$0.042&2.254$\pm$0.107&1.577&0.731&0.680$\pm$0.029&-44.07$\pm$3.71&149.81$\pm$5.18&1.45&4.66\\
7410.943&0.407&1.435$\pm$0.005&1.646$\pm$0.015&1.128$\pm$0.042&3.438$\pm$0.014&1.691&1.754&1.217$\pm$0.161&-2.48$\pm$1.11&184.00$\pm$3.60&1.46&3.97\\
7410.966&0.410&1.377$\pm$0.001&1.575$\pm$0.012&1.119$\pm$0.060&3.371$\pm$0.002&1.701&1.671&1.074$\pm$0.156&-5.15$\pm$1.15&198.74$\pm$3.55&1.41&4.41\\
7411.949&0.557&1.191$\pm$0.002&1.398$\pm$0.012&1.010$\pm$0.043&2.543$\pm$0.002&1.373&1.169&0.651$\pm$0.061&-30.03$\pm$1.42&223.67$\pm$3.41&1.38&5.49\\
7411.972&0.560&1.191$\pm$0.005&1.387$\pm$0.011&1.009$\pm$0.054&2.402$\pm$0.026&1.360&1.055&0.540$\pm$0.020&-29.31$\pm$1.72&210.92$\pm$4.04&1.37&6.26\\
7412.935&0.703&1.192$\pm$0.009&1.379$\pm$0.019&0.974$\pm$0.058&2.401$\pm$0.001&1.438&0.963&0.673$\pm$0.051&-22.58$\pm$1.11&222.14$\pm$2.91&1.42&5.02\\
7412.992&0.712&1.182$\pm$0.010&1.362$\pm$0.014&0.972$\pm$0.063&2.549$\pm$0.004&1.417&1.133&0.668$\pm$0.061&-16.26$\pm$1.12&230.58$\pm$3.16&1.40&5.37\\
7414.931&0.000&1.079$\pm$0.002&1.278$\pm$0.014&0.926$\pm$0.045&2.349$\pm$0.046&1.435&0.937&0.588$\pm$0.042&-34.89$\pm$1.21&178.37$\pm$2.27&1.38&5.62\\
7415.016&0.013&1.064$\pm$0.001&1.265$\pm$0.016&0.929$\pm$0.054&2.405$\pm$0.028&1.456&0.964&0.565$\pm$0.016&-37.66$\pm$1.45&188.40$\pm$2.75&1.36&5.99\\
7415.937&0.150&1.185$\pm$0.003&1.413$\pm$0.021&1.004$\pm$0.058&2.736$\pm$0.020&1.449&1.297&0.799$\pm$0.069&-23.83$\pm$0.81&166.11$\pm$1.92&1.41&4.82\\
7416.012&0.161&1.211$\pm$0.004&1.421$\pm$0.018&1.029$\pm$0.064&2.745$\pm$0.026&1.463&1.295&0.728$\pm$0.017&-20.11$\pm$0.87&161.36$\pm$2.23&1.38&5.30\\
7416.939&0.299&1.249$\pm$0.009&1.433$\pm$0.016&1.030$\pm$0.065&2.837$\pm$0.020&1.740&1.107&0.736$\pm$0.030&-31.95$\pm$1.32&181.80$\pm$2.62&1.39&5.42\\
7416.962&0.302&1.257$\pm$0.010&1.437$\pm$0.013&1.025$\pm$0.055&2.807$\pm$0.035&1.687&1.138&0.757$\pm$0.050&-31.46$\pm$1.38&177.39$\pm$2.68&1.40&5.22\\
7417.971&0.452&1.163$\pm$0.006&1.375$\pm$0.016&0.989$\pm$0.074&1.847$\pm$0.076&1.311&0.574&0.469$\pm$0.017&-66.82$\pm$3.74&152.54$\pm$5.38&1.39&5.54\\
7417.995&0.456&1.175$\pm$0.008&1.357$\pm$0.018&0.990$\pm$0.075&1.836$\pm$0.089&1.342&0.538&0.475$\pm$0.007&-75.26$\pm$3.98&150.25$\pm$6.03&1.37&5.43\\
7418.941&0.597&1.662$\pm$0.004&1.986$\pm$0.020&1.342$\pm$0.079&7.895$\pm$0.066&2.262&5.255&2.804$\pm$0.014&31.76$\pm$0.66&307.25$\pm$1.88&1.48&3.96\\
7418.985&0.603&1.578$\pm$0.002&1.897$\pm$0.018&1.296$\pm$0.064&7.759$\pm$0.037&2.185&5.083&2.367$\pm$0.015&31.45$\pm$1.13&281.71$\pm$2.40&1.46&4.61\\
7419.012&0.607&1.555$\pm$0.003&1.854$\pm$0.019&1.284$\pm$0.072&7.440$\pm$0.024&2.058&4.861&2.368$\pm$0.010&31.12$\pm$1.35&271.29$\pm$2.57&1.44&4.42\\
\enddata
\tablecomments{H$_{\alpha}$$_{n}$ and H$_{\alpha}$$_{b}$ represent the narrow and broad components of the H$_{\alpha}$ line, respectively.}
\end{deluxetable*}
\end{longrotatetable}
\bibliography{sample631}{}
\bibliographystyle{aasjournal}


\end{document}